# Ratchet propagation of a magnetic domain wall in a single magnetic wire with quantum interference

Akinobu Yamaguchi[1,2], Tomoaki Kishimoto[1] and Hideki Miyajima[1]

[1] Department of Physics, Keio University, 3-14-1, Hiyoshi, Yokohama 223-8522, Japan
[2] PRESTO, JST, 4-1-8, Honcho, Kawaguchi, Saitama 332-0012, Japan

CORRESPONDENCE should be addressed to A. Y. (e-mail: yamaguch@phys.keio.ac.jp) and H. M. (email: miyajima@phys.keio.ac.jp)

**Abstract**

Quantum interference incorporating spatially asymmetric potential profiles is realized experimentally to manipulate a magnetic domain wall (DW) into a single multilayered wire whose spacer has a thickness gradient for generating asymmetrical interlayer exchange coupling from side to side. We demonstrate experimentally how to guide a DW in a micron-scale ferromagnetic wire without reflection symmetry of the interlayer exchange coupling. This is the ratcheting of a DW in a form of ratchet potential using quantum interference. The experimental results can be described well by numerical simulations considering spatially asymmetric potential profiles due to quantum interference.

A ratchet is a machine part consisting of a toothed wheel or bar that allows movement in only one direction. By analogy, a microscopic ratchet system consists of a periodic potential that lacks reflection symmetry. The non-equilibrium dynamics of a particle in a ratchet potential has long been studied as a fundamental problem in physics[1]. Ratchet systems have attracted greater attention in recent years because of their newly found relevance to diverse areas of physics, chemistry and biology. A single particle confined in an asymmetric potential is expected to demonstrate a ratchet effect by drifting along the easy ratchet direction when subjected to non-equilibrium fluctuations[1-3]. This well-known effect should occur even if the particle is replaced with a magnetic domain wall (DW) that separates regions of opposite magnetization.

A simple way to achieve ratchet propagation of a DW in a ferromagnetic/nonmagnetic/ferromagnetic (F/N/F) multilayer wire with quantum interference is to asymmetrically modulate the exchange coupling along the wire, as shown in Fig. 1. The exchange coupling between ferromagnetic films separated by a nonmagnetic metal layer does not decrease monotonically as the spacer-layer thickness increases, but instead exhibits an oscillatory dependence with a succession of antiferromagnetic and ferromagnetic coupling in a variety of thickness ranges[4-10]. The oscillatory behaviour of the exchange coupling is ascribed to exchange interactions propagated by the spin-dependent confinement of electrons, leading to quantum interference as a function of the spacer-layer thickness[4-12].

In this study, we report the first experimental results on ratchet propagation of a DW in a single multilayered wire whose spacer layer has a thickness gradient which generates coupling that is asymmetrical from side to side.

We began our experiment with a three-layered wire of $Fe_{19}Ni_{81}/Au/Fe_{19}Ni_{81}$. It was grown on thermally oxidized silicon substrates using electron-beam lithography, ultrahigh vacuum electron-beam evaporation to create a wedge-shaped Au spacer layer and the lift-off method. The sample was 1-μm wide, and the ferromagnetic layers were 50-nm and 10-nm thick, respectively. The Au wedge was deposited at room temperature and ranged in thickness from 0 to 12 nm, with a gradient of $2\times10^{-6}$ directed along the length of the wire. Electrodes were made by patterning Cr (5 nm)/Au (100 nm) films. The voltage contacts were 5 mm apart, and the thickness of the Au wedge between them ranged from 1 to 11 nm, as shown in the schematic in Fig. 1. One end of the wire was tapered to a sharp point to prevent nucleation of the DW, thus ensuring that the DW was injected only from the blunt edge of the wire[13, 14].

Magnetoresistance (MR) depends on the relative orientation of the magnetic moments of ferromagnetic layers, giving rise to giant magnetoresistance (GMR) in the case of alternating ferromagnetic and nonmagnetic layers. MR ratio is usually defined as $(R_{AF}-R_F)/R_F$, where $R_{AF}$ and $R_F$ are the resistances for antiferromagnetic (AF) and ferromagnetic (F) alignment of the magnetic moments, respectively. The magnitudes of the GMR and switching field $H_S$ are directly related to the exchange coupling[6-12]. The DW's position in the wire can be identified by the change in electrical resistance described by the above relation. As the GMR ratio of the wires is proportional to the length of the switching-layer magnetization, we can evaluate the DW's position using the resistance measurements[15].

To confirm that the exchange coupling oscillates with spacer-layer thickness, we measured the variation in the current-in-plane GMR with Au spacer-layer thickness using shadow-masked junctions[6-10], in which thickness of Au is kept constant across the wire.

With application of a magnetic field along the wire, the resistance was measured using a standard four-point DC technique at room temperature. Representative results are shown in Fig. 2. Typical examples of MR loops for ferromagnetic- and antiferromagnetic-coupling are shown in the insets of Fig. 2, although it was hard to distinguish between the ferromagnetic- and antiferromagnetic-coupling only from just the shape of the field dependence of MR. The Au dependence of $H_S$ are shown in Fig. 2.

Note that $H_S$ does not decrease monotonically with the thickness of Au, but it does exhibit an oscillatory dependence. The oscillation-period $D_0$ of $H_S$ is estimated at $D_0 = 2.5$ nm. To further evaluate the generality of our findings, we performed curve fitting for the data in Fig. 2. We reproduced qualitatively the oscillatory behaviour in Fig. 2 using the relation

$$H_{\mathrm{S}} = \frac{H_0}{d}\cos\left(\frac{2\pi d}{D_0}\right) + \mathrm{const.}, \qquad (1)$$

where $H_0$ denotes the amplitude parameter we reproduced the experimental results, and $d$ is the thickness of Au. The observed oscillatory behaviours of the GMR ratio and $H_{\mathrm{S}}$ are attributed to quantum interference[11, 12]. Figure 2 indicates that at least two antiferromagnetic coupling positions exist in the wire, the spacer layer that has a thickness gradient from 1 nm to 11 nm in 5 mm.

To test these predictions experimentally, we measured MR of the wire without the reflection symmetry of the spacer layer (Fig. 1). Typical results for the slope dependence of MR loop are shown in Figs. 3a and b. A wire with a constant Au-layer thickness of 12 nm exhibits the normal magnetization reversal process[14, 15], independent of the DW injection direction (not shown here). These results indicate that almost no defects prevent DW propagation toward the opposite edge. However, an unusual magnetization switching process is observed in the multilayered wire with a thickness gradient. Two-step resistance decreases are observed during the magnetization reversal (Fig. 3b), whereas the DW nucleates from the blunt edge and moves rapidly to the other edge (Fig. 3a). This indicates that the asymmetric wedge structure plays a crucial role in DW displacement. When we measured the GMR loops of multilayers with a lower thickness gradient, a single step appeared during magnetization reversal (not shown here). This means that the DW stopped at a position where the antiferromagnetic coupling is stable, and the energy barrier is between the local minimum and maximum. These results demonstrate remarkable ratchet propagation of a single DW.

Using the well–established one-dimensional (1D) model[16–19], we can now understand ratchet propagation of a DW. In the present model, DW retains its profile as it moves. The wall dynamics can be described by just two parameters, its position $X$ and its conjugate momentum, defined as $2M_S\phi/\gamma$, where $\phi$ is the tilt angle of the wall magnetization out of the plane of the wire (Fig. 4a), $M_S$ is the saturation magnetization and $\gamma$ is the gyromagnetic factor. The 1D model has been described by two simultaneous equations[16-19],

$$\left(1+\alpha^2\right)\dot{X} = \alpha\gamma\Delta\left[H_{ext} + H_{ex}(X)\right] + \gamma\Delta H_K \sin\phi\cos\phi, \qquad (2)$$

and

$$\left(1+\alpha^2\right)\dot{\phi} = \gamma\left[H_{ext} + H_{ex}(X)\right] - \alpha\gamma H_K \sin\phi\cos\phi, \qquad (3)$$

where $\Delta$ is the width of the DW, $\alpha$ is the Gilbert damping parameter, $H_{ext}$ is the effective field including the external magnetic field, $H_{ex}(X)$ is the effective field derived from the exchange coupling and $H_K$ is the anisotropy field derived from the shape and crystalline anisotropy energy.

In the derivation of Eqs. (2) and (3), the magnetic field along the traveling direction of the DW gives the potential energy gradient, yielding the force[16-19]. Here, assuming that $H_{ex}(X)$ is given by Eq. (1), we obtain the potential $V(x)$ for the DW induced by exchange coupling[19],

$$V(x) = \int M(x) \cdot H(x) \, dx = \int dx \tanh \frac{x-X}{\Delta} \cdot \frac{H_0 \cos \kappa x}{\kappa x}, \quad (4)$$

where $d = \kappa x$ with inclination coefficient $\kappa$. Force is obtained by[19]

$$F = \frac{\partial V(x)}{\partial X} = \frac{1}{\Delta} \int dx \frac{1}{\cosh^2 \frac{x-X}{\Delta}} \cdot \frac{H_0 \cos \kappa x}{\kappa x} \simeq \int_{X-\frac{\Delta}{2}}^{X+\frac{\Delta}{2}} dx \frac{H_0 \cos \kappa x}{\kappa x}. \quad (5)$$

In the limit that the DW's width is far smaller than the wavelength of the exchange coupling along the wire, we take $\kappa \Delta \ll 1$, and we immediately see from Eq. (5) that

$$F \simeq \frac{H_0 \cdot \Delta \cos \kappa X}{\kappa X}. \quad (6)$$

This result allows us to replace the effective exchange field $H_{ex}(X)$ with Eq. (6), corresponding to the switching field $H_S$ given by Eq. (1).

To examine the validity of Eqs. (2), (3) and (6), we carried out a simulation on a wire having the same dimensions as the present wire. Assuming the spatial distribution of the exchange coupling in the wire shown in Fig. 2b, we calculated the effective exchange field $H_{ex}(X)$ as a function of the DW position (inset of Fig. 4b). The fit obtained using Eqs. (1) and (6) matches the experimental data.

The 1D model describes well the essential physical mechanisms responsible for the ratchet motion of DW with space inversion asymmetry of the interlayer exchange coupling, as shown in Fig. 4. The behaviour agrees well with the experimental results. For a negative direction of DW displacement in the asymmetrical exchange field, both $X$ and $\phi$ are roughly independent of the external field $H_{ext}$ (Figs. 4a and b). However, for a positive direction of DW displacement in the same asymmetrical exchange field, the DW movement distance $X$ and $\phi$ show a clear staircase structure (Fig. 4c) and rotation (Fig. 4d), respectively. The simulation results qualitatively reproduced the experimental results. This indicates that exchange coupling without reflection symmetry can produce a ratchet potential with respect to DW propagation.

The present experiments provide clear evidence for ratchet propagation of a single DW in a ferromagnetic wire with space inversion asymmetry of the interlayer exchange coupling. This phenomenon is quite different from the reversible steps that result from the pinning derived from a defect. In our experiments, the DW displacement was controlled by modulation of the quantum potential well. The experimental results agree quite well with the predictions of the model. This suggests that the model captures the essential physics of the observed ratchet propagation of a DW in a ferromagnetic wire with quantum interference. This ratchet behaviour will provide a crucial tool for controlling and manipulating the motion of a DW and magnetization reversal. Finally, we stress that our findings could be useful in tailoring the exchange coupling by an appropriate choice of thickness gradient of the nonmagnetic layer.

We acknowledge Mr. Matsumoto's technical support of the shutter system. We thank Prof. G. Tatara and Prof. Y. Nakatani for fruitful discussions. We appreciate Prof. M. Nakasako's support for the usage of his atomic force microscope. The present study was partly supported by the Sumitomo Foundation, MEXT Grants-in-Aid for Scientific Research in a Priority Area, the JST PRESTO and a JSPS Grant-in-Aid for Scientific Research.

# [References]

**[Figure Captions]**

Figure 1 Sample properties, concept principle and experiment. Schematic of the sample structure and the cross section along the longitudinal axis of the wire. One end of the wire is tapered to a sharp point to prevent nucleation of a magnetic domain wall (DW), which ensures the DW is injected only from the blunt edge. Interlayer exchange coupling has oscillatory properties as a function of the nonmagnetic layer thickness. The magnetic DW, which separates regions of opposite magnetization, is displaced in the wire without reflection symmetry of the exchange coupling.

Figure 2 Switching field as a function of nonmagnetic layer thickness. Insets show MR measurements of a wire with $t_N$ = 3.7 and 5.5 nm. Switching field $H_S$ of pinned ferromagnetic layer as a function of Au thickness. Black solid line shows the fitting curve for $H_S$ given by Eq. (1).

Figure 3 MR loops of a 5 mm long wire with space inversion asymmetry of the exchange coupling. During magnetization reversal in the pinned ferromagnetic layer F2, normal and clear two–step magnetization reversal processes are exhibited through the GMR effect in (a) and (b), respectively. Anomalous behaviour of magnetization reversal depends on the direction of DW

displacement.

Figure 4 Ratchet dynamics of a domain wall in a wire without reflection symmetry of exchange coupling, calculated with a 1D model. Gilbert damping, gyromagnetic factor and saturation magnetization are $\alpha = 0.01$, $\gamma = 1.76$ MHz/Oe and 10800 Oe, respectively. The DW position $X$ and tilt angle $\phi$ of the wall magnetization out of the plane of the wire are shown in the inset of (a). Assuming that the interlayer exchange field $H_{ex}(X)$ varies as indicated in Eq. (1), we carried out simulations. Calculated dependence of $H_{ex}(X)$ on a position $X$ in the wire is shown in the inset of (b). For a DW injected from the right blunt edge and propagating in the negative $x$-direction, the dependences of $X$ and $\phi$ on the external magnetic field along the longitudinal axis of the wire are shown in (a) and (b), respectively. As shown in (b), the magnetization precesses only during the DW's motion. The external magnetic field dependences of $X$ and $\phi$ for DW propagation in the opposite direction are shown in (c) and (d), respectively.

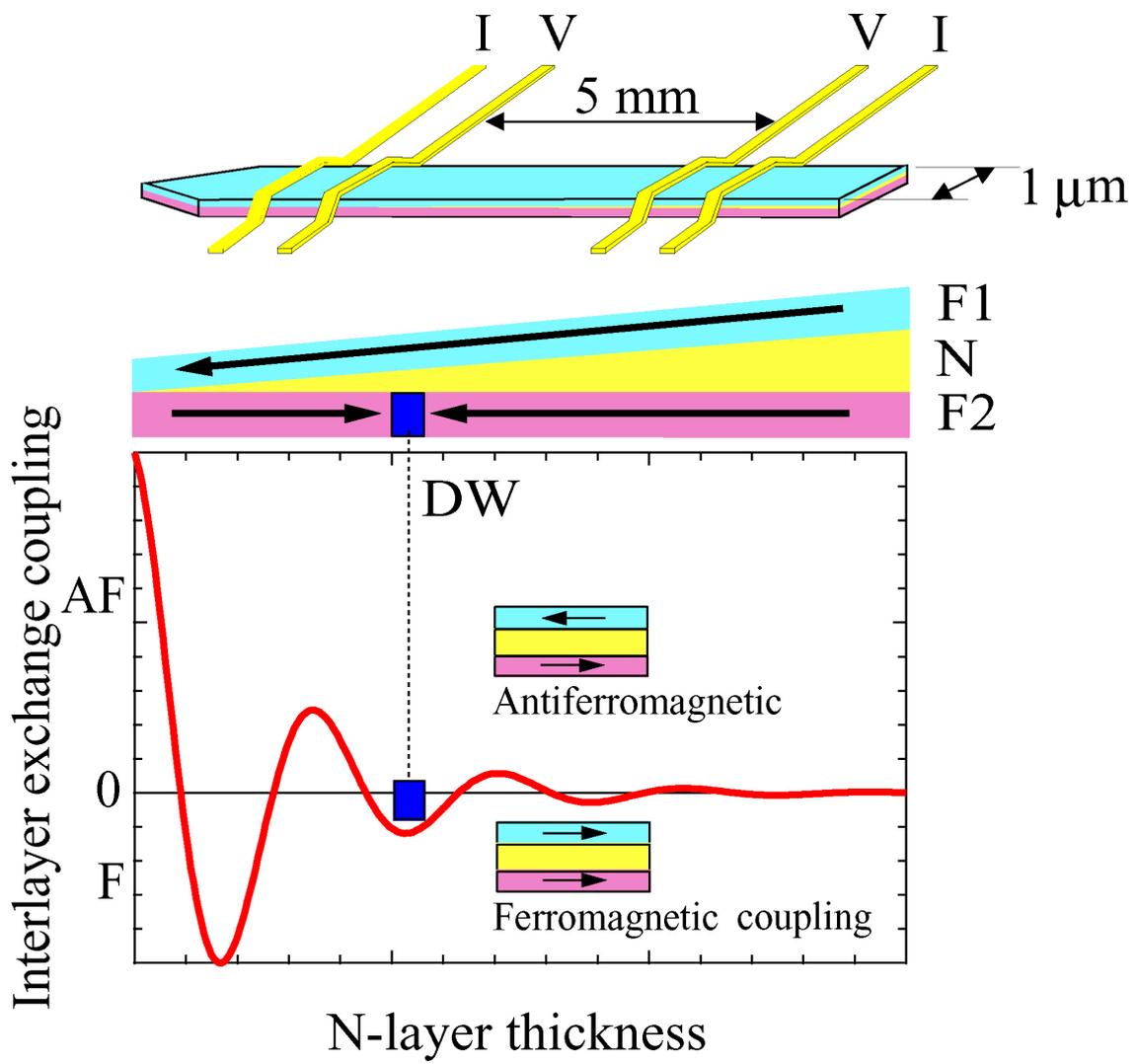

Fig. 1

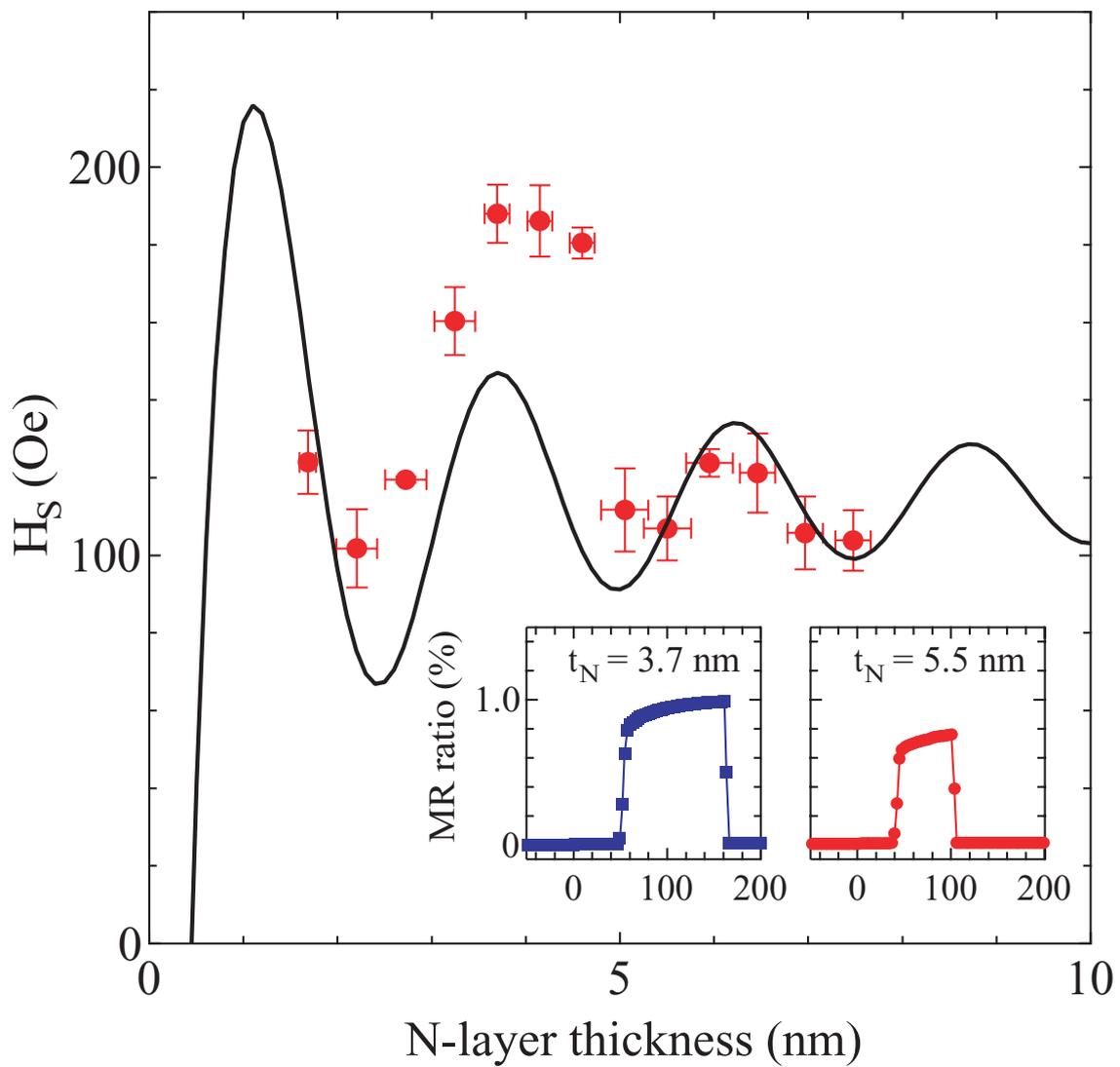

Fig. 2

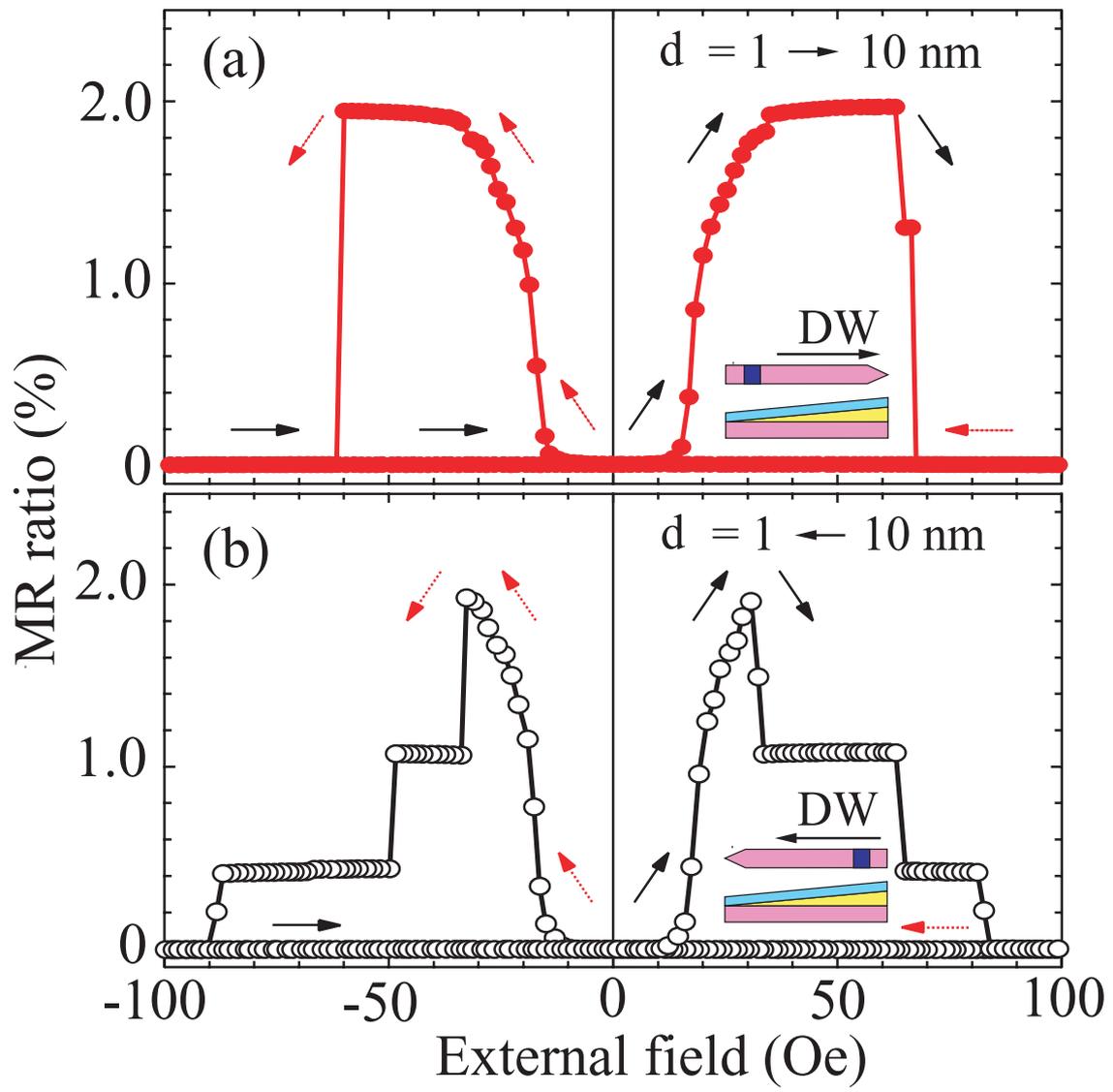

Fig. 3

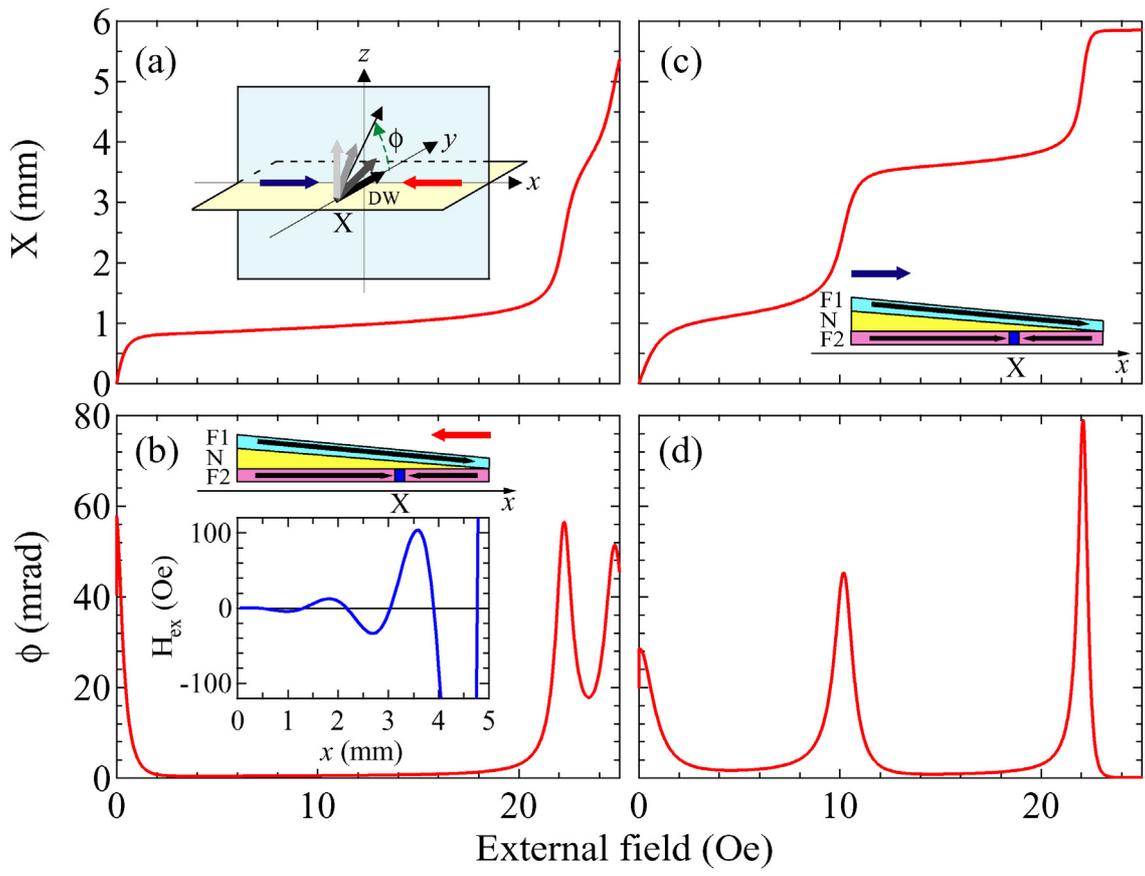

Fig. 4